# Effects of Different Levels of Self-Representation on Spatial Awareness, Self-Presence and Spatial Presence during Virtual Locomotion

Jingbo Zhao    Zhetao Wang    Yaojun Wang

College of Information and Electrical Engineering

China Agricultural University

{zhao.jingbo, wangyaojun}@cau.edu.cn

*Abstract*—Recently, there has been growing interest in investigating the effects of self-representation on user experience and perception in virtual environments. However, few studies investigated the effects of levels of body representation (full-body, lower-body and viewpoint) on locomotion experience in terms of spatial awareness, self-presence and spatial presence during virtual locomotion. Understanding such effects is essential for building new virtual locomotion systems with better locomotion experience. In the present study, we first built a walking-in-place (WIP) virtual locomotion system that can represent users using avatars at three levels (full-body, lower-body and viewpoint) and is capable of rendering walking animations during in-place walking of a user. We then conducted a virtual locomotion experiment using three levels of representation to investigate the effects of body representation on spatial awareness, self-presence and spatial presence during virtual locomotion. Experimental results showed that the full-body representation provided better virtual locomotion experience in these three factors compared to that of the lower-body representation and the viewpoint representation. The lower-body representation also provided better experience than the viewpoint representation. These results suggest that self-representation of users in virtual environments using a full-body avatar is critical for providing better locomotion experience. Using full-body avatars for self-representation of users should be considered when building new virtual locomotion systems and applications.

## I. Introduction

Many commercial VR applications and games still have not adopted the practice of using avatars for self-representation of users in virtual environments. Representation of a user using a viewpoint (without an avatar) is still commonly used by many VR applications. Under such condition, VR users are floating above the ground in virtual environments. This can cause spatial awareness difficulties and can also result in a lowered sense of presence in virtual environments [1]. For example, perception of one's exact position in virtual environments is difficult as users are floating in virtual environments, and they cannot see their virtual legs as a visual reference on the virtual ground. It can be difficult for users to determine their exact location in virtual environments under this condition. Other spatial awareness difficulties include the perception of one's height, distance to virtual objects and sizes of virtual objects, etc. Presence refers to the sense of "being there" [2]. It has many dimensions that include self-presence and spatial presence, etc. Self-presence is the sensation that one is convinced that their virtual body is their actual self [3] and spatial presence refers to the feeling of being physically present in virtual environments [4]. The lack of avatars for self-representation of users may result in a lowered sense of self-presence and spatial presence. Different levels of body representation may also affect these two factors. On the other hand, self-representation of users using a full-body avatar, a lower-body avatar or a viewpoint is commonly used by many first-person perspective (FPP) games. Recent FPP games tend to use the full-body representation for self-representation of users while earlier games often use the lower-body representation and the viewpoint representation to represent users. The settings of FPP games inspired us to include these three levels of body representation in the present study as VR applications may also use the full-body representation and the lower-body representation for user self-representation.

Although there has been growing interest in investigating the effects of self-representation of users on user experience in virtual environments [4]–[16], the investigation of the effects of different levels of body representations on user experience, particularly on virtual locomotion experience, was largely ignored by previous research. An investigation is required to examine the effects of different levels of body representation on virtual locomotion experience.

To this end, we developed a WIP virtual locomotion system that can represent a user at three different levels and is capable of rendering walking animation while a user is performing in-place walking. This system was implemented based on a previous work [17]. The reason to implement a WIP system for this research was because of its low-cost and there is no requirement for large tracking space. WIP systems may also be the most applicable solution for ordinary VR users to perform virtual locomotion due to these two reasons. Thus, understanding the effects of body representation is important for WIP systems. In addition, this system design would also enable other researchers to reproduce this work using VR devices that include a VR headset and a Kinect v2 sensor. Based on this system, we conducted a virtual locomotion experiment to investigate the effects of three levels of body representation on spatial awareness, self-presence and spatial presence.

Our results show that using the full-body representation provided better virtual locomotion experience in terms of spatial awareness, self-presence and spatial presence compared to the lower-body representation and the viewpoint representation and that the lower-body representation also provided better experience than that of the viewpoint representation. This result is important as it suggests that using a full-body avatar for self-representation of users provides better experience than other two representations. VR engineers and designers should consider using a full-body avatar for self-representation of users when building new virtual locomotion systems and applications.

The rest of the paper is organized as follows: Section II presents related work; Section III describes the hardware and software of the VR system and the implementation of the WIP locomotion system with three different levels of body representation; Section IV presents the experiment and the results; Section V draws the conclusions of the study.

## II. RELATED WORK

Understanding the effects of the egocentric view of self-representation using different levels of body representations is important for building better locomotion systems. Previous studies that are closely related to the present research examined the effects of self-representation in terms of avatar personification [10], avatar appearance [13] and levels of body representation under conditions including hands only, hands and feet without an avatar and various tracking points with an avatar [4] using various tasks in virtual environments.

Waltermate et al. [10] examined the effects of personalized avatars on body ownership, presence and emotional response. This study included three levels of avatar appearance, which were generic avatars, generic scanned avatars and individualized scanned avatars. Participants were asked to perform a series of body movements under each condition and assess their experience in VR. Their results showed that individualized avatars significantly increased body ownership, presence and dominance compared to generic counterparts.

Wirth et al. [13] studied user experience and gait variability under conditions with a combination of different levels of avatar appearance, camera perspectives and the contexts of virtual environments while participants were walking on a treadmill. Definitions of body representation in their study were no avatar, point cloud representation, silhouette representation, and full-body male and female avatar representation, as experimental conditions. Their results showed that different levels of avatar appearance have significant effects on gait, user experience and presence. Participants preferred full-body avatar representation to others.

Gonçalves et al. [4] used different levels of body representation that included virtual hands only (H), virtual hands and feet without an avatar (HF), full-body avatar with three-point tracking to enable head and hands movement (3P), three-point tracking with simulated walking (3PS), five-point tracking that added the control of virtual feet (5P) and six-point tracking that allowed the control of the avatar's hip movement (6P). Their experimental tasks required participants to find hidden objects in the virtual scene and place those objects in a travel bag. Their results showed that hands only (H) had significantly better-experienced realism than hands and feet (HF) but had no significant effects on other factors. They also found that there were no significant differences between hands only (H) and three-point tracking with a full-body avatar (3P), and between hands and feet (HF) and five-point tracking with a full-body avatar (5P). In addition, this study found that increasing the tracking points from five (5P) to six (6P) did significantly improve response and embodiment but there were no differences between three-point tracking (3P), three-point tracking with simulated walking (3PS) and five-point tracking (5P).

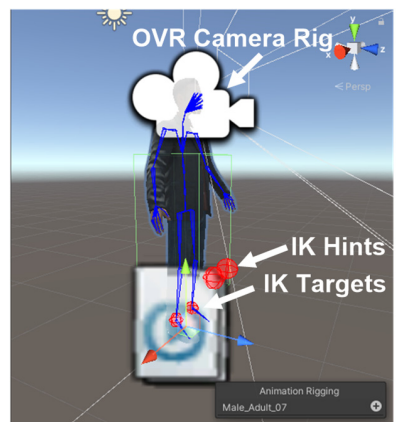

Fig. 1. IK targets, IK hints and OVR camera rig setup in Unity.

Through literature review, we found that existing studies had their own definition of body representation in their experiments. The effects of body representation were examined based on their selection of factors using different VR tasks. No previous research has investigated the effects of different levels of body representation, including the full-body representation, the lower-body representation (legs only) and the viewpoint representation (no body representation), on spatial awareness, self-presence and spatial presence during virtual locomotion. This is worth investigating in our research.

## III. METHODS

### A. Hardware and Software

The experimental hardware of the WIP virtual locomotion system included a desktop computer equipped with an Intel i5 11400F CPU, 16 GB memory and an Nvidia RTX 3060 graphics card with 12 GB graphics memory, an Oculus Quest 2 VR headset and a Kinect v2 sensor. The headset was wirelessly connected to the desktop computer through a Wifi 6 router and was used to present the virtual environment. The Kinect sensor was used to capture users' body movements in real-time. The experimental software application, hosted on the desktop computer, was developed using Unity 2019.02. The experimental software application presented the virtual scene for locomotion and rendered three different levels of user body representation during in-place walking in virtual environments.

### B. Implementation of a WIP Virtual Locomotion System with Three Different Levels of Body Representation

A kinematic approach was adopted for this WIP system to generate leg animation during in-place walking using foot tracking data captured by a Kinect v2 sensor. Specifically, inverse kinematic (IK) targets, provided by Unity's animation rigging package, were attached to an avatar's feet and IK hints were placed in front of the knees empirically to complete the basic setup (See Fig. 1). The derived kinematic equations were then used to guide the IK targets to perform cyclic motions based on the foot tracking data captured by a Kinect v2 sensor [17]. Since the IK targets were attached to the feet of an avatar, these IK targets guide the virtual feet to perform reciprocating walking motion while a user was moving their feet up and down during in-place walking.

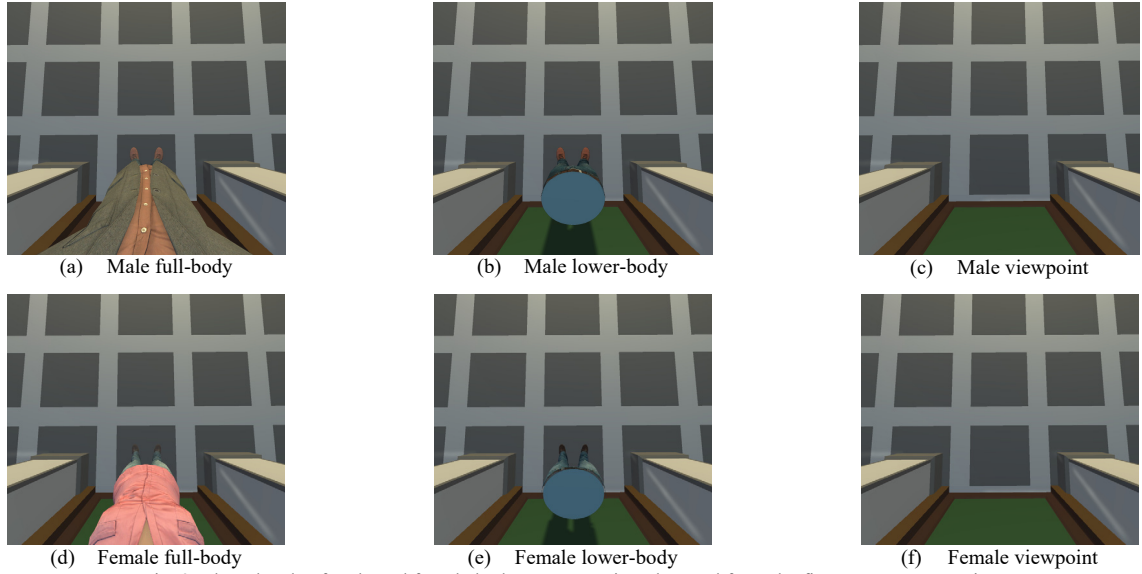

Fig. 2. Three levels of male and female body representation observed from the first-person perspective.

Table I. Spatial awareness, self-presence and spatial presence questionnaire

| | | |
|---|---|---|
| **Spatial Awareness** | 1. | My avatar helps me perceive my distance to objects in virtual environments. |
| | 2. | My avatar helps me perceive the sizes of objects in virtual environments. |
| | 3. | My avatar helps me perceive my height in virtual environments. |
| | 4. | My avatar helps me keep balance while walking in virtual environments. |
| | 5. | My avatar helps me locate/orient myself in virtual environments. |
| **Self-Presence** | 1. | To what extent was the avatar an extension of yourself? |
| | 2. | To what extent did you feel if something happened to the avatar, it felt like it was happening to you? |
| | 3. | To what extent did you feel that the avatar's body was your own body? |
| | 4. | To what extent did you feel that the avatar was you? |
| | 5. | How much did the avatar's movements correspond with your actions? |
| **Spatial presence** | 1. | To what extent did you feel like you were really inside the virtual space? |
| | 2. | To what extent did you feel surrounded by the virtual space? |
| | 3. | To what extent did you feel like you really visited the virtual space? |
| | 4. | To what extent did you feel that the virtual space seemed like the real world? |
| | 5. | To what extent did you feel like you could reach out and touch the objects in the virtual space? |

For upper body movements of the full-body representation, we generated arm animation by directly mapping the tracked angles of shoulder, elbow and wrist joints of a user, estimated by a Unity package (Azure Kinect Examples for Unity) using the captured Kinect sensor data, to the corresponding joints of an avatar. Consequently, when users moved their arms in the physical space, they could see the movements of their virtual arms in the virtual space.

The avatar models to represent users' virtual bodies were selected from the Microsoft Rocketbox avatar library [18]. We selected a male model and a female model so that male and female participants had the opportunity to select the avatar with the corresponding gender to represent themselves during the virtual walking experiment. The full-body representations of males and females are shown in Fig. 2a and Fig. 2d. For the lower-body representation and the viewpoint representation, we edited the textures of the avatar models and painted related parts of the textures transparent. Specifically, to represent a user's lower body (legs only), we removed the region of a texture that corresponded to the upper body of the avatar and painted the removed region transparent. The lower body part of the texture was kept unchanged. To prevent participants from seeing the inside surface of the lower body of an avatar model after the region of the texture corresponding to the upper body was removed, we covered the top of the waist of avatar models with a blue 3-D elliptical plane. By applying the edited texture to the avatar model and adding the blue 3-D elliptical plane, only the lower body was visible (see Fig. 2b and Fig. 2e). For the viewpoint representation (see Fig. 2c and Fig. 2f), the textures of avatar models were painted completely transparent so that users were not able to see the avatars while they were in virtual environments. The OVR camera rig of the Oculus VR headset was placed at the eye position of the avatar models in all three conditions (see Fig. 1). When users wore the VR headset and lowered their head to look down, they were able to see their virtual body representation.

### C. Questionnaire

The questionnaire used for the study was divided into three sections: spatial awareness, self-presence and spatial presence.

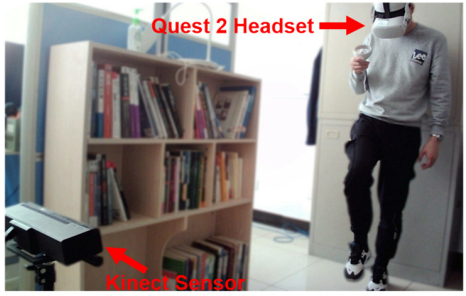

Fig. 3. Experimental Setup. Demonstration by a researcher, performing in-place walking with movements captured by a Kinect v2 sensor.

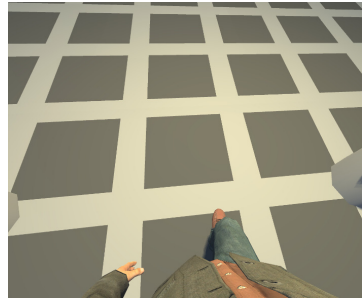

Fig. 4. First-person perspective view (male, full-body) during in-place walking.

We designed a spatial awareness questionnaire, shown in Table I, based on the spatial awareness problems discussed in [1]. This questionnaire enabled participants to evaluate their perception of egocentric distance to virtual objects in the virtual scene, the sizes of virtual objects, their height in virtual environments, balance and their position and orientation in virtual environments. Participants were asked to rate each item using a seven-point Likert scale (from strongly disagree to strongly agree). This section of questionnaire is exploratory and would require additional validation in future studies.

The self-presence section and spatial presence section were adapted from previous studies [19], [20]. These questionnaires were also rated using a seven-point Likert scale (from extremely small to extremely large). Although the original questionnaire was designed based on a five-point Likert scale [19], changing the scale format from five-point to seven-point should not affect the data characteristics according to previous research [21].

## IV. EXPERIMENT

### A. Goal

In this experiment, we aimed to investigate the effects of different levels of self-representation on spatial awareness, self-presence and spatial presence during virtual locomotion.

### B. Design

The virtual scene that presented an indoor office area for users to walk was a free package, called "Snaps Prototype | Office", acquired from the Unity asset store. This indoor office area has a corridor of thirty meters, which made it suited to conduct our virtual locomotion experiment. The task asked a participant to walk along the thirty-meter corridor. A collision volume was placed at the end of the corridor so that when participants entered this volume, their task is completed. We adopted a within-subject design for the experiment. There were three conditions: (a) Full-body; (b) Lower-body (legs only) and (c) Viewpoint (no avatar). In (a) Full-body, participants were able to see their arms, upper trunk and legs. Their virtual arms moved accordingly as they moved their physical arms. In (b) Lower-body, participants were able to see their virtual legs but not their virtual trunk and arms. In (c) Viewpoint, participants were floating in the virtual environment as the texture of the avatar was painted completely transparent. We counter-balanced the order of levels of body representation for each participant to control for order effects. For each experimental session, there were three blocks that corresponded to Full-body, Lower-body and Viewpoint. In each block, there were four walking trials. The first trial was a practice trial, and the remaining three trials were experimental trials. In each trial, a participant was asked to walk for thirty meters along the corridor of the office area in the virtual scene until they reached the bounding volume. After finishing each block, the participant was asked to fill the questionnaire, as shown in Table I. The participants were then asked to complete the same experiment in another block.

### C. Participants

We invited twenty-four students (ten males and fourteen females, age: 18-25 and height: 156 cm-180 cm) to participate in our experiment. Informed consents were provided to participants and were completed before experiments.

### D. Procedure

Before experiments, a Kinect v2 sensor was mounted on a tripod approximately 60 cm high from the ground, with the camera view capturing a participant (see Fig. 3, the experimental setup was demonstrated by a researcher capturing himself). During an experiment, a participant was introduced to the procedure of the experiment and completed the informed consent. The participant then stood approximately 1.8 m in front of the Kinect sensor wearing the Quest 2 VR headset. The researcher then asked the participant to level their heads and reset the view of the VR headset so that the view of the participant was aligned to the eye position of the avatar model. Subsequently, a 3-s calibration capture was conducted, which recorded the participant's foot position to estimate the reference height of the floor plane to calculate foot height based on captured foot position data. After the calibration procedure finished and when the participant was ready, the researcher initiated the trial with a 3-s countdown timer, followed by a "start" prompt, to inform the participant to start in-place walking. Participants were allowed to freely observe the virtual environment and their self-representation avatar during walking (see Fig. 4). They continued their walking until they reached the collision volume to complete the walking trial. Participants were asked to fill the questionnaire once they finished a complete block. Participants were then introduced to the next block using another body representation. Walking trials were conducted following the procedure described above. The same process was repeated until participants completed all blocks.

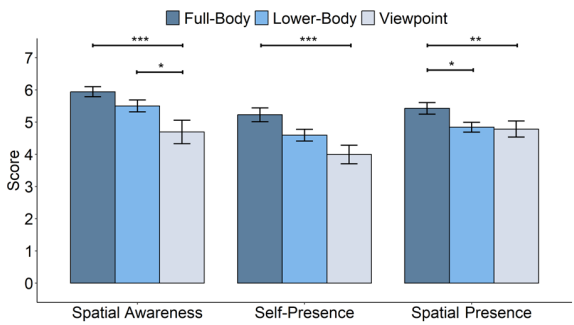

Fig. 5. Average scores of spatial awareness, self-presence and spatial presence calculated across items in each section of the questionnaire for all participants. Error bars denote the standard error of the mean. Significance bars denote the pairs with significant differences.

*E. Results*

Statistical analyses were performed using R 4.2.1. As in [22], we adopted the linear mixed-effects analyses to analyze our experimental data. The independent factors of the analyses were the three levels of representation, and the dependent factors were the scores of spatial awareness, self-presence and spatial presence. Initially, gender was included as an independent factor, but the effect was weak so it was excluded from the model. This could be attributed to the limited number of samples that were not sufficient to test the gender effect on dependent factors. Effect sizes were reported using partial eta squared $\eta_p^2$ (estimated from repeated-measures ANOVA analyses of the same form as the linear mixed-effects Models analyses). To perform statistical analyses, we calculated the average scores of these three factors across items in each section of the questionnaire for all participants (weighting the items in each section can be considered for further exploration). The average scores and the standard error of the mean are shown in Fig. 5. Cronbach's alpha is calculated for each body representation using the average scores from three sections of the questionnaire. Results indicated good internal consistency on full-body ($\alpha = 0.775$) and viewpoint ($\alpha = 0.869$) and acceptable consistency on lower-body ($\alpha = 0.684$). Linear mixed-effects analyses then were applied to analyze the experimental data. Results showed that different levels of body representation have significant effects on spatial awareness ($F(2, 46) = 8.999, p < 0.001, \eta_p^2 = 0.28$), self-presence ($F(2, 46) = 9.191, p < 0.001, \eta_p^2 = 0.29$) and spatial presence ($F(2, 46) = 6.072, p = 0.005, \eta_p^2 = 0.21$). Post-hoc pairwise comparisons were conducted using Tukey's range tests. As shown in Fig. 5, the pairs with significant differences are indicated by significance bars. By examining Fig. 5, we conclude that Full-body has significantly higher scores for spatial awareness, self-presence and spatial presence compared to Viewpoint. This result shows that Full-body can indeed enhance one's spatial awareness, self-presence and spatial presence in virtual environments compared to Viewpoint. Although there was a significant difference between Lower-body and Viewpoint in terms of spatial awareness, there was no significant difference between Full-body and Lower-body for this factor. This indicates that having a self-representation using either a full-body avatar or a lower-body avatar is better than no avatar and having an avatar with either a full-body avatar or a lower-body avatar as a visual reference is important for spatial awareness in VR. For self-presence, there was no significant difference between Full-body and Lower-body and between Lower-body and Viewpoint. But average scores were the highest for Full-body, and the lowest for Viewpoint while lower-body was in the middle. In terms of spatial presence, there was a significant difference between Full-body and Lower-body, indicating that the fully-body representation enhances one's spatial presence.

## V. CONCLUSION

In this study, we showed that the full-body representation for virtual locomotion had better experience for users in terms of spatial awareness, self-presence and spatial presence compared to the lower-body representation and the viewpoint representation. The lower-body representation was also better than the viewpoint representation as shown by experimental results. To conclude, this research provided empirical evidence that the full-body representation provides better experience compared to two other body representation methods during virtual walking. Thus, using full-body avatars to represent users should be considered when designing new virtual locomotion systems that intend to incorporate user self-representation. Finally, the present research used a WIP system to conduct the virtual locomotion experiment, further experiments using other virtual locomotion techniques, including real walking, redirected walking [23] and mechanical repositioning [24], also need to be conducted to validate the results of this research on those techniques.